\documentclass[prl,aps,twocolumn,superscriptaddress]{revtex4}
\usepackage{epsfig}

\begin{document}

\title{Spontaneous coherence of indirect excitons in a trap}

\author{A.A.~High}
\affiliation{Department of Physics, University of California at San Diego, La Jolla, CA 92093-0319, USA}
\author{J.R.~Leonard}
\affiliation{Department of Physics, University of California at San Diego, La Jolla, CA 92093-0319, USA}
\author{M.~Remeika}
\affiliation{Department of Physics, University of California at San Diego, La Jolla, CA 92093-0319, USA}
\author{L.V.~Butov}
\affiliation{Department of Physics, University of California at San Diego, La Jolla, CA 92093-0319, USA}
\author{M.~Hanson}
\affiliation{Materials Department, University of California at Santa Barbara, Santa Barbara, CA 93106-5050, USA}
\author{A.C.~Gossard}
\affiliation{Materials Department, University of California at Santa Barbara, Santa Barbara, CA 93106-5050, USA}
\date{\today}

\begin{abstract}
\noindent We report on the emergence of spontaneous coherence in a gas of indirect excitons in an electrostatic trap. At low temperatures, the exciton coherence length becomes much larger than the thermal de~Broglie wavelength and reaches the size of the exciton cloud in the trap.
\end{abstract}

\maketitle

Potential traps are an effective tool for studies of cold atomic gases. They allow the realization and control of atomic Bose-Einstein condensates, see \cite{Cornell02, Ketterle02} for review. Condensation in momentum space is equivalent to the emergence of spontaneous coherence of matter waves \cite{Penrose56}. Spontaneous coherence is an intensively studied feature of atomic condensates \cite{Cornell02, Ketterle02}. In this paper, we report on studies of spontaneous coherence in a cold gas of indirect excitons in an electrostatic potential trap.

Excitons are hydrogen-like electron-hole pairs at low densities \cite{Keldysh68} and Cooper-pair-like electron-hole pairs at high densities \cite{Keldysh65}. The bosonic nature of excitons allows for condensation in momentum space at low temperatures, below the temperature of quantum degeneracy. For a typical range of parameters, the temperature of quantum degeneracy in an exciton gas is in the range of a few Kelvin. Although the temperature of the semiconductor crystal lattice can be lowered well below 1 K in He-refrigerators, lowering the temperature of the exciton gas to even a few Kelvin is challenging \cite{Tikhodeev98, Jang06}. Due to recombination, excitons have a finite lifetime which is too short to allow cooling to low temperatures in regular semiconductors. In order to create a cold exciton gas with temperature close to the lattice temperature, the exciton lifetime should be large compared to the exciton cooling time. Besides this, the realization of a cold and dense exciton gas requires an excitonic state to be the ground state and have lower energy than the electron-hole liquid \cite{Keldysh86}.

These requirements can be fulfilled in a gas of indirect excitons. An indirect exciton in coupled quantum wells (CQW) is a bound state of an electron and a hole in separate wells (Fig. 1a). The spatial separation allows one to control the overlap of electron and hole wavefunctions and engineer structures with lifetimes of indirect excitons orders of magnitude longer than those of regular excitons \cite{Lozovik76, Fukuzawa90}. In our experiments, indirect excitons are created in a GaAs/AlGaAs CQW structure (Fig.~1a). Long lifetimes of the indirect excitons allow them to cool to low temperatures within about $0.1\,\text{K}$ of the lattice temperature, which can be lowered to about $50\,\text{mK}$ in an optical dilution refrigerator \cite{Butov01}. This allows the realization of a cold exciton gas with temperature well below the temperature of quantum degeneracy $T_{\text{dB}} = 2\pi \hbar^2 n / (m g k_{\text{B}})$ (in the studied CQW, excitons have the mass $m = 0.22 m_0$, spin degeneracy $g = 4$, and $T_{\text{dB}} \approx 3\,\text{K}$ for the density per spin $n / g = 10^{10}\,\text{cm}^{-2}$).

\begin{figure}
\includegraphics[width=5.5cm]{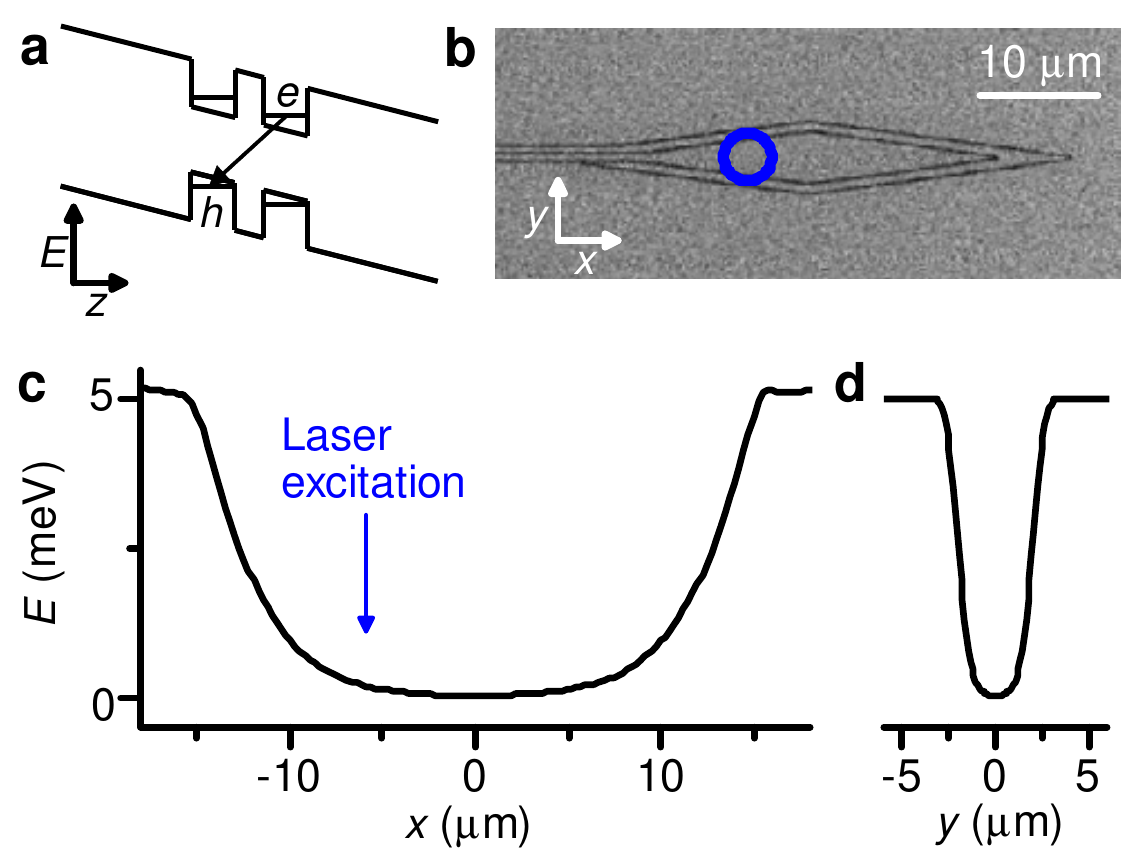}
\caption{(a) CQW band diagram. e, electron; h, hole. Indirect excitons are formed by electrons and holes confined in separated layers. (b) SEM image of electrodes forming the diamond trap: a diamond-shaped electrode is surrounded by a thin wire electrode followed by an outer plane electrode. (c,d) Simulation of exciton energy profile through the trap center along $x$ (c) and $y$ (d) for $V_{diamond} = -2.5$ V,  $V_{wire} = -2$ V, and $V_{plane} = -2$ V. The position of the laser excitation spot is indicated by the circle in (b) and by the arrow in (c).}
\end{figure}

Due to the spatial separation of the electron and hole, indirect excitons have a built-in dipole moment $ed$, where $d$ is close to the distance between the QW centers, which allows their energy to be controlled by voltage: an electric field $F_z$ perpendicular to the QW plane results in the exciton energy shift $E = e d F_z$ ~\cite{Miller85}. This gives an opportunity to create in-plane potential landscapes for excitons $E(x,y) = e d F_z(x,y)$. Advantages of electrostatically created potential landscapes include the opportunity to realize the desired potential profile as well as the ability to control it in-situ, on a time scale shorter than the exciton lifetime. For instance, switching off the confining potential, like in atomic time-of-flight experiments, or modulating its depth, like in atomic collective mode experiments \cite{Cornell02, Ketterle02} can be realized by modulating the electrode voltage. Excitons were studied in various electrostatic potential landscapes: ramps \cite{Hagn95, Gartner06}, lattices \cite{Zimmermann98, Krauss04, Hammack06, Remeika09}, circuit devices \cite{High08}, and traps \cite{Hammack06, Huber98, Chen06, High09prl, Gorbunov06}.

In this work, an electrostatic trap for indirect excitons is realized using a diamond-shaped electrode (Fig. 1b). The diamond trap creates a parabolic-like confining potential for excitons \cite{High09prl}. The CQW structure is grown by MBE. An $n^+$-GaAs layer with $n_{Si}=10^{18}$ cm$^{-3}$ serves as a homogeneous bottom electrode. The top electrodes on the surface of the structure are fabricated via e-beam lithography by depositing a semitransparent layer of Ti (2 nm) and Pt (8 nm). The device includes a $3.5 \times 30 \mu$m diamond electrode, a 600 nm wide 'wire' electrode which surrounds the diamond, and 'outer plane' electrode (Fig. 1b) \cite{High09prl}. Two 8 nm GaAs QWs separated by a 4 nm Al$_{0.33}$Ga$_{0.67}$As barrier are positioned 100 nm above the $n^+$-GaAs layer within an undoped 1 $\mu$m thick Al$_{0.33}$Ga$_{0.67}$As layer. Positioning the CQW closer to the homogeneous electrode suppresses the in-plane electric field \cite{Hammack06}, which otherwise can lead to exciton dissociation. The excitons are photoexcited by a 633 nm HeNe laser. The excitation beam is focused to a spot about $5 \mu$m in diameter on a side of the trap (Fig. 1b,c). This excitation scheme allows the photoexcited excitons to cool down further when they travel towards the trap center, thus facilitating the realization of a cold and dense exciton gas in the trap (cooling of excitons away from the laser excitation spot also leads to the realization of a cold exciton gas in the inner ring in exciton emission pattern \cite{Butov02, Ivanov06} and in laser-induced traps \cite{Hammack06a}).

\begin{figure}
\includegraphics[width=8.5cm]{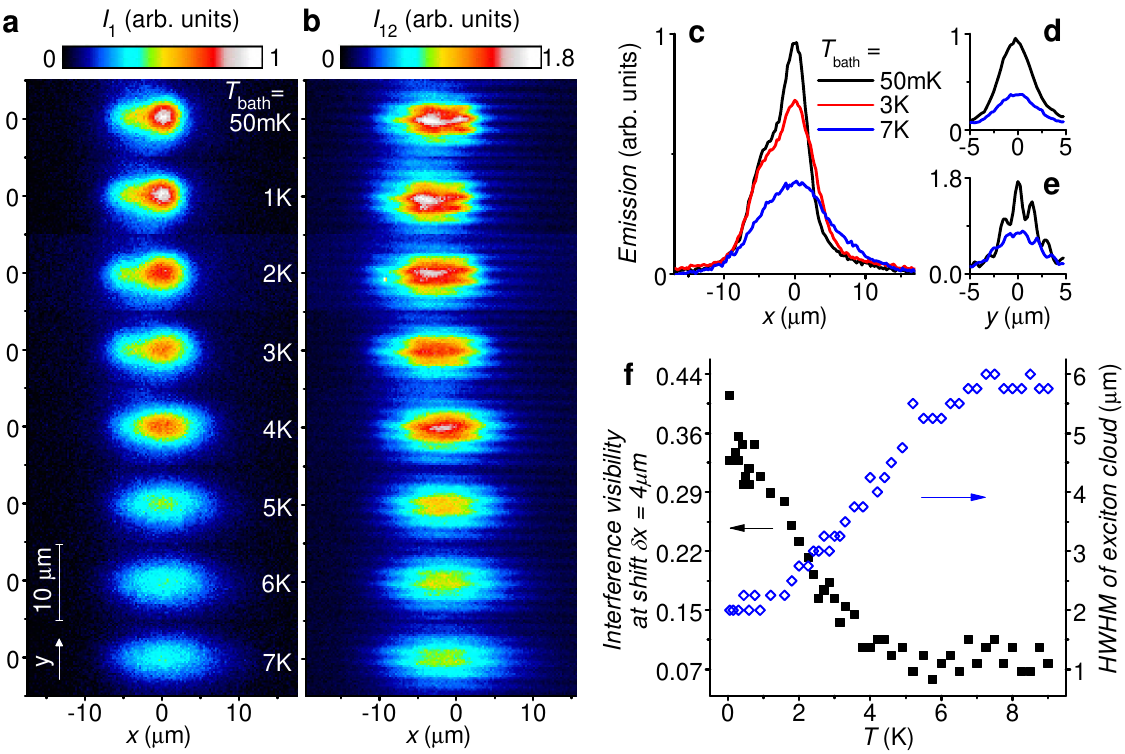}
\caption{(a,b) Emission pattern (a) and interference pattern at shift $\delta x = 4\,\mu$m (b) for temperatures ranging from 50 mK to 8 K. (c,d) Spatial profiles of the emission pattern in (a) along $x$ at $y=0$ (c) and along $y$ at $x=0$ (d) for $T_{\text{bath}} = 50\,\text{mK}$ (black), $3\,\text{K}$ (red), and $7\,\text{K}$ (blue). (e) Spatial profiles of the interference patterns in (b) along $y$ at the peak of exciton emission ($x=0$) for $T_{\text{bath}} = 50\,\text{mK}$ (black) and $7\,\text{K}$ (blue). (f) Amplitude of the interference fringes $A_{interf}$ at shift $\delta x = 4\,\mu$m averaged from $0<x<1.5 \mu$m (black squares) and half-width at half-maximum (HWHM) of the exciton emission pattern along $x$ (blue diamonds) vs. temperature. For all data $P_{ex} = 1.9\,\mu$W.}
\end{figure}

The pattern of the first-order coherence function $g_1(\delta x)$ is measured by shift-interferometry: the emission images produced by arm 1 and 2 of the Mach-Zehnder interferometer are shifted with respect to each other to measure the interference between the emission of excitons separated by $\delta x$. A similar method was used in the studies of spontaneous coherence in a gas of indirect excitons in exciton rings \cite{High11}. The emission beam is made parallel by an objective inside the optical dilution refrigerator and lenses. The emission is split between arm 1 and arm 2 of the interferometer. The path lengths of arm 1 and arm 2 are set equal. The interfering emission images produced by arm 1 and 2 of the interferometer are shifted relative to each other along $x$ to measure the interference between the emission of excitons, which are laterally separated by $\delta x$. After the interferometer, the emission is filtered by an interference filter of linewidth $\pm 5\,\text{nm}$ adjusted to the emission wavelength of indirect excitons $\lambda = 800\,\text{nm}$. The filtered signal is focused to produce an image, which is recorded by a liquid-nitrogen cooled CCD. We measure emission intensity $I_1$ for arm 1 open, intensity $I_2$ for arm 2 open, and intensity $I_{12}$ for both arms open, and calculate $I_{\text{interf}} = (I_{12} - I_1 - I_2) / (2 \sqrt{I_1 I_2})$. The period of the interference fringes is set by a tilt angle between the image planes of the two arms and $g_1(\delta x)$ is given by the amplitude of the interference fringes $A_{\text{interf}}$ \cite{Milonni88} as detailed below.

Figure 2 presents the temperature dependence of exciton emission and interference patterns. At high temperatures, the exciton cloud spreads over the trap resulting in a broad spatial profile of the exciton emission (Fig. 2a,c,f). With lowering temperature, the width of the emission pattern of the exciton cloud in the trap reduces indicating exciton accumulation at the trap center (Fig. 2a,c,f). This is consistent with the reduction of the thermal spreading of excitons over the trap. Note that studies of atoms in traps also reveal the accumulation of atoms at the trap center with lowering temperature \cite{Cornell02,Ketterle02}.

Figure 2b presents the temperature dependence of the pattern of interference fringes. The corresponding temperature dependence of the amplitude of the interference fringes $A_{\text{interf}}$  averaged from $0<x<1.5 \mu$m is shown in Fig. 2f. As detailed below, $A_{\text{interf}}$ presents the coherence degree of excitons. Figure 2f shows that the exciton accumulation at the trap center is accompanied by the enhancement of the coherence degree of excitons.

\begin{figure}
\includegraphics[width=5cm]{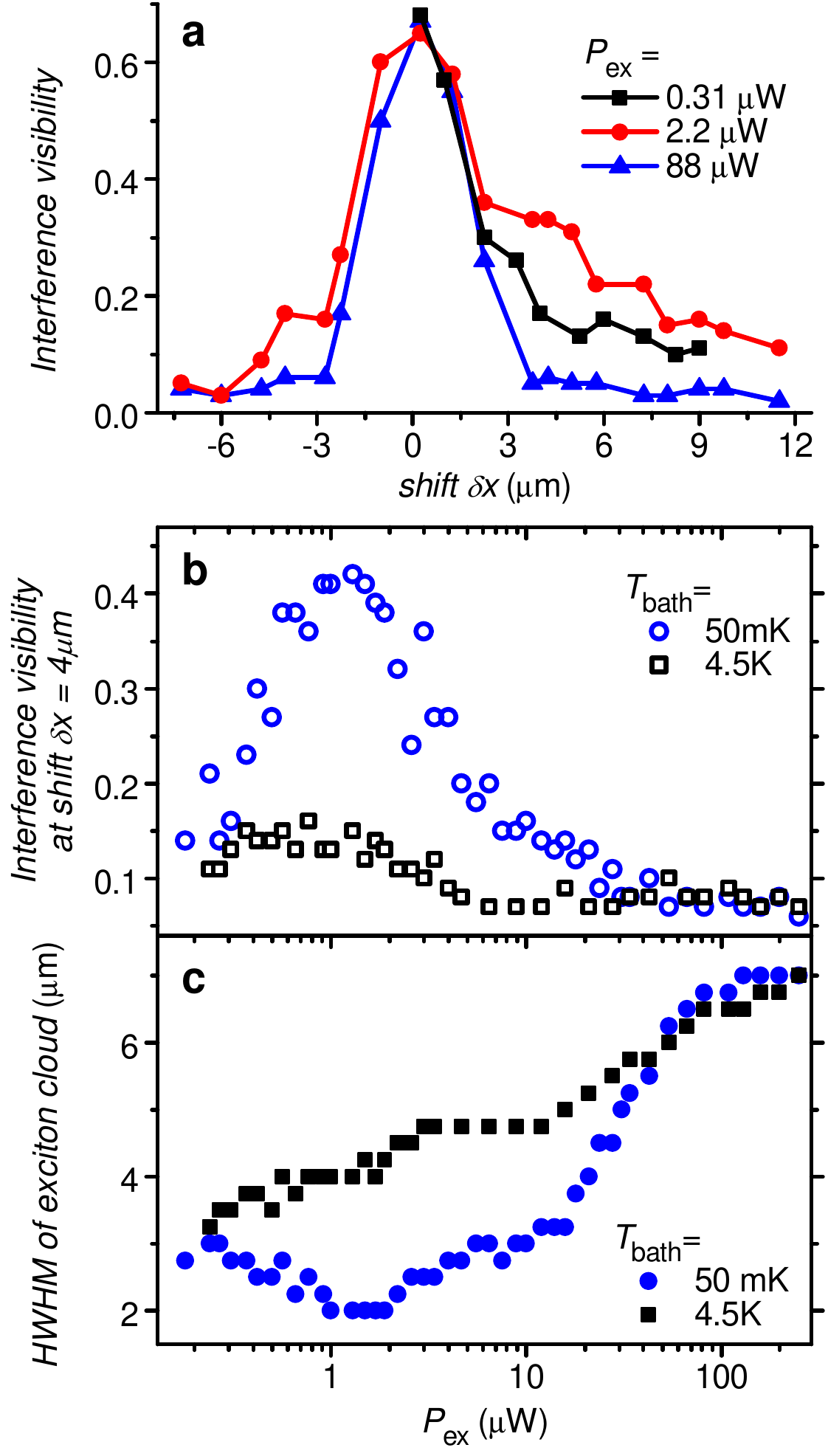}
\caption{ (a) Amplitude of the interference fringes $A_{\text{interf}}(\delta x)$ for excitons in the trap for excitation density $P_{ex}=0.31\,\mu$W (black squares), $2.2\,\mu$W (red circles), and $88\,\mu$W (blue triangles) at $T_{\text{bath}} = 50\,\text{mK}$. (b,c) Amplitude of the interference fringes $A_{interf}$ at shift $\delta x = 4\,\mu$m averaged from $0<x<1.5 \mu$m (b) and HWHM of the exciton emission pattern along $x$ (c) vs. $P_{ex}$ for $T_{\text{bath}} = 50\,\text{mK}$ (blue circles) and $4.5\,\text{K}$ (black squares).}
\end{figure}

Figure 3a presents the amplitude of the interference fringes $A_{\text{interf}}(\delta x)$ for different densities. The exciton temperature is higher in the excitation spot \cite{Ivanov06, Hammack06a}. This is consistent with low values of $A_{\text{interf}}$ at negative $\delta x$, which correspond to the interference between the emission of a hot exciton gas in the excitation spot and exciton gas in the trap center (Fig. 3a). We will consider positive $\delta x$, which correspond to the interference between the emission of an exciton gas in the trap center and exciton gas at positive $x$ further away from the hot laser excitation spot.

Figure 3c presents the density dependence of the width of the exciton emission pattern along $x$. At high temperature $T = 4.5$ K, the width of the emission pattern of the exciton cloud monotonically increases with density (Fig. 3c). This is consistent with screening of the potential landscape in the trap by indirect excitons due to the repulsive exciton-exciton interaction \cite{High09prl}. However, at low temperature $T = 50$ mK, the width of the emission pattern of the exciton cloud nonmonotonically changes with density: the increase of density first leads to the reduction of the cloud width, indicating the exciton accumulation at the trap center, while at higher densities, an increase of the cloud width with density is observed (Fig. 3c).

Figure 3b presents the density dependence of the amplitude of the interference fringes $A_{\text{interf}}$. At high temperature $T = 4.5$ K, the coherence degree is low for all densities (Fig. 3b). However, at low temperature $T = 50$ mK, the coherence degree nonmonotonically changes with density: the increase of density first leads to a strong enhancement of $A_{\text{interf}}$, followed by its reduction (Fig. 3c). Note that maximum $A_{\text{interf}}$ corresponds to minimum width of the emission pattern of the exciton cloud in the trap (Fig. 3b,c).

\begin{figure}
\centering
\includegraphics[width=5.5cm]{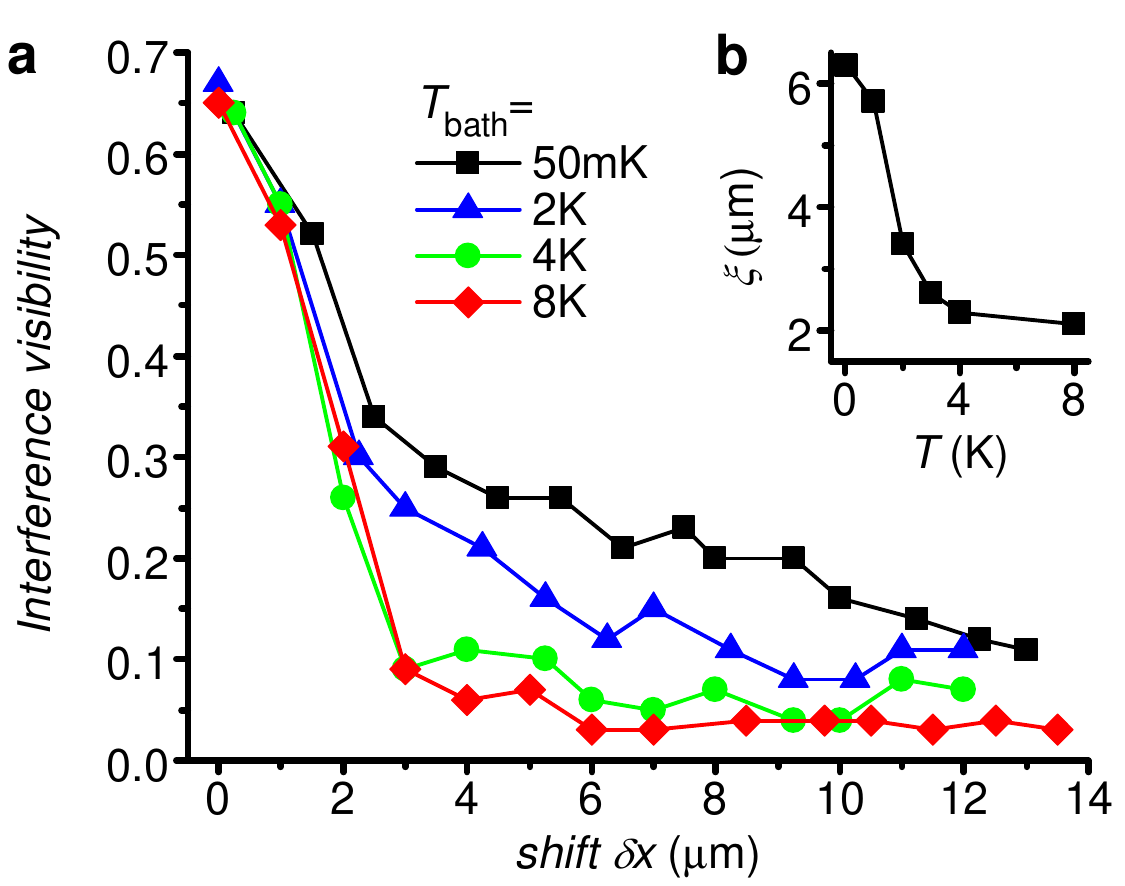}
\caption{(a) $A_{\text{interf}}(\delta x)$ for excitons in the trap for $T_{\text{bath}} = 50\,\text{mK}$ (black squares), $2\,\text{K}$ (blue triangles), $4\,\text{K}$ (green circles), and $8\,\text{K}$ (red diamonds). (b) $\xi$ evaluated as $\delta x$ at which the interference visibility drops $e$ times. For all data $P_{ex} = 1.9\,\mu$W.}
\end{figure}

Figure 4 presents the amplitude of the interference fringes $A_{\text{interf}}(\delta x)$ for different temperatures. At high temperatures $T \gtrsim 4 K$, $A_{\text{interf}}$ quickly drops with $\delta x$. As discussed below, this behavior is expected for a classical gas. However, extended spontaneous coherence is observed at low temperatures (Fig. 4). The spatial extension of $A_{\text{interf}}(\delta x)$ can be characterized by a coherence length $\xi$. Here, to consider all data on equal footing, we evaluate $\xi$ as $\delta x$ at which the interference visibility drops $e$ times. The temperature dependence of $\xi$ is presented in Fig. 4b. A strong enhancement of the exciton coherence length is observed at low temperatures. While at high temperatures $\xi$ is considerably smaller than the exciton cloud width, at low temperatures the entire exciton cloud becomes coherent (Figs. 2f and 4).

The data are discussed below. In the reported experiments, the laser excitation energy exceeds the exciton energy by about 400 meV and the laser excitation spot is spatially separated from the interfering excitons for positive $\delta x$. Therefore studied coherence in the exciton gas is spontaneous coherence; it is not induced by coherence of the laser excitation. [Note that for negative $\delta x$, the interfering excitons spatially overlap with the laser excitation spot and exciton coherence is suppressed (Fig. 3a), confirming that exciton coherence studied in this work is not induced by the laser excitation.]

Coherence of exciton gas is imprinted on coherence of exciton emission, which is described by the first-order coherence function $g_1(\delta x)$. In turn, this function is given by the amplitude of the interference fringes $A_{\text{interf}}(\delta x)$ in `the ideal experiment' with perfect spatial resolution. In real experiments, the measured $A_{\text{interf}}(\delta x)$ is given by the convolution of $g_1(\delta x)$ with the point-spread function (PSF) of the optical system used in the experiment \cite{High11,Fogler08}. The PSF width corresponds to the spatial resolution of the optical system.

At high temperatures, the amplitude of interference fringes quickly drops with $\delta x$ and the width of $A_{\text{interf}}(\delta x)$ corresponds to the PSF. This behavior is characteristic for a classical gas, where $g_1(\delta x)$ drops within the thermal de~Broglie wavelength $\lambda_{dB} = \sqrt{\frac{2 \pi \hbar^2}{m T}}$, which is about $0.5 \mu$m at $0.1\,\text{K}$, below the PSF width. At low temperatures, extended exciton coherence with the coherence length much larger than in a classical gas is observed (Fig. 4). At the lowest temperatures, the observed coherence length in the exciton gas in the trap exceeds $\lambda_{dB}=0.5\, \mu$m at 0.1 K by an order of magnitude and the entire exciton cloud in the trap becomes coherent (Figs. 2f and 4).

Figure 4a illustrates why $\delta x = 4 \mu$m is selected for presenting coherence degree of excitons in Figs. 2 and 3. The shift $\delta x = 4 \mu$m is chosen to exceed both $\lambda_{dB}$ and the PSF width. At such $\delta x$, only weak coherence given by the PSF value at $\delta x = 4 \mu$m can be observed for a classical gas. Higher $A_{\text{interf}}$ exceeding such background level reveal the enhanced coherence degree of excitons.

In conclusion, we report on the emergence of spontaneous coherence in a gas of indirect excitons in a trap. At low temperatures, the exciton coherence length becomes much larger than the thermal de~Broglie wavelength and reaches the size of the exciton cloud in the trap.

We thank Michael Fogler for discussions. This work was supported by
ARO grant W911NF-08-1-0341. The development of spectroscopy in a dilution refrigerator was also supported by the DOE award DE-FG02-07ER46449 and NSF grant 0907349.

\end{document}